\newlength\smallfigwidth
\documentclass[aps,prb, 10pt,twocolumn,floatfix,showpacs,amsmath,amssymb,superscriptaddress] {revtex4-1}

\usepackage{color}
\usepackage{amsmath}
\usepackage{graphicx}
\usepackage{graphics}
\usepackage{pdftexcmds}
\usepackage{ifpdf}
\usepackage{amssymb}
\usepackage{esint}
\usepackage{comment}

\makeatletter
\@ifundefined{textcolor}{}
{%
 \definecolor{BLACK}{gray}{0}
 \definecolor{WHITE}{gray}{1}
 \definecolor{RED}{rgb}{1,0,0}
 \definecolor{GREEN}{rgb}{0,1,0}
 \definecolor{BLUE}{rgb}{0,0,1}
 \definecolor{CYAN}{cmyk}{1,0,0,0}
 \definecolor{MAGENTA}{cmyk}{0,1,0,0}
 \definecolor{YELLOW}{cmyk}{0,0,1,0}
 }

\allowdisplaybreaks
\makeatother

\def \Sp {\mathbf{S}}
\def \Dp {\mathbf{D}}

\def \ev {\mathbf{e}}
\def \Bv {\mathbf{B}}
\def \Mv {\mathbf{M}}
\def \rv {\mathbf{r}}

\def \be {\begin{equation}}
\def \ee {\end{equation}}
\def \bea {\begin{eqnarray}}
\def \eea {\end{eqnarray}}
\allowdisplaybreaks
\makeatother

\begin{document}

\title{Composite Spin Crystal Phase in Antiferromagnetic Chiral Magnets}

 \author{S.A. Osorio}
 \affiliation{IFLP, UNLP, CONICET, Facultad de Ciencias Exactas, C.C. 67, 1900 La Plata, Argentina and Departamento de Física, FCE, UNLP, La Plata, Argentina}
 \author{H.D. Rosales}
 \affiliation{IFLP, UNLP, CONICET, Facultad de Ciencias Exactas, C.C. 67, 1900 La Plata, Argentina and Departamento de Física, FCE, UNLP, La Plata, Argentina}
 \author{M.B. Sturla}
 \affiliation{IFLP, UNLP, CONICET, Facultad de Ciencias Exactas, C.C. 67, 1900 La Plata, Argentina and Departamento de Física, FCE, UNLP, La Plata, Argentina}
 \author{D.C. Cabra}
 \affiliation{IFLP, UNLP, CONICET, Facultad de Ciencias Exactas, C.C. 67, 1900 La Plata, Argentina and Departamento de Física, FCE, UNLP, La Plata, Argentina}
 \affiliation{Abdus Salam International Centre for Theoretical Physics, Associate Scheme, Strada Costiera 11, 34151, Trieste, Italy}

\date{\today}

\begin{abstract}

We study the classical antiferromagnetic Heisenberg model on the triangular lattice with Dzyaloshinskii-Moriya interactions in a magnetic field. We focus in particular in the emergence of a composite spin crystal phase, dubbed antiferromagnetic skyrmion lattice, that was recently observed in [Phys. Rev. B 92, 214439 (2015)] for intermediate fields. This complex phase can be made up from three inter-penetrated skyrmion lattices, one for each sub-lattice of the original triangular one. Following these recent numerical results, in this paper we explicitly construct the low-energy effective action that reproduces the correct phenomenology and could serve as a starting point to study the coupling to charge carriers, lattice vibrations, structural disorder and transport phenomena.
\end{abstract}

\maketitle

%%%%%%%%%%%%%%%%%%%%%%%%%%%%%%%%%%%%%%%%%%%%%%%%%%%%%%%%%%%%%%%%%%%%%%%%%%%%%%%%
\section{Introduction}
%%%%%%%%%%%%%%%%%%%%%%%%%%%%%%%%%%%%%%%%%%%%%%%%%%%%%%%%%%%%%%%%%%%%%%%%%%%%%%%%

Antiferromagnets have been the focus of an enormous amount of work, mainly since the
suggestion that they could be at the origin of the pairing mechanism in High TC
superconductors \cite{Anderson_87}.

On the other hand, in some chiral magnets such as
MnSi\cite{MBJ_09,ITB_76,IA_84,GDM_09,LHP_95,PRP_04,JGM_13},
Fe$_{1-x}$Co$_x$Si\cite{BVR_83,GDM_07,OTT_05},
FeGe\cite{LBF_89,UNH_08,YKO_11,WBS_11}, and
Mn$_{1-x}$Fe$_x$Ge\cite{SYH_13}, a new kind of complex magnetic
structure has been observed. This new phase, known as Skyrmion
crystal, observed in some region of temperatures and magnetic fields, consists
in a periodic arrangement of topologically protected magnetic textures
that resemble the one proposed by Skyrme\cite{Skyrme_62}.

The existence of these topological nano-sized spin structures in
condensed matter, called magnetic skyrmions, are well know since long
time ago. They appear in different systems like
liquid-crystals\cite{WM_89}, quantum-Hall ferromagnets\cite{SKK_93},
Bose condensate\cite{Ho_98}, etc.

The potential technological applications of this phase of
chiral magnets are numerous. Among others, the possibility of driving the motion of
the magnetic skyrmions with ultra-low current densities, an anomalous
Hall effect, and the observed multi-ferroic behavior makes these systems
particularly interesting for applications to processing devices and
information storage, in particular to race-track memory
devices\cite{SCR_13,TMZ_14}. On the other hand, the existence
of high frequency periodic excitations of the skyrmion lattice phase,
makes them promising candidates for nano-scale microwave
resonators \cite{Schwarze_2015}.

The underlying mechanism responsible for this structure seems to be an
anti-symmetric spin orbit interaction, known as Dzyaloshinskii-Moriya
interaction (DM)\cite{Dzyaloshinsky_58,Moriya_60}. In generic
non-centro-symmetric magnetic crystals a DM interaction can stabilize
a skyrmion crystal phase.
The existence of these topologically protected structures in chiral
magnets was theoretically predicted
in \cite{BY_89,BH_94,RBP_06}. Later on, Yi {\it et al}. \cite{YON_09} have
shown by Monte Carlo simulations that a classical ferromagnetic spin
system with DM interaction supports, in a given region of the
parameter space, skyrmion lattice structures. Han {\it et al}. \cite{HZY_10}
have proven that a non-linear sigma model plus a continuous version of
the DM interaction in a magnetic field, proposed as the low energy
Hamiltonian of these chiral magnets, reproduces the observed
phenomenology.

In a recent work \cite{Rosales_15}, a detailed Monte Carlo simulation
has shown the existence of an exotic magnetic phase on a triangular antiferromagnetic lattice,
in the presence of a DM interaction and for a certain window in the external magnetic field.
This exotic phase, named AF-SkX, consists of a periodic arrangement of sets of spins which can be reinterpreted as a three-flavor 
interpenetrated skyrmion lattice. Such phase arises in a frustrated simple antiferromagnetic model which exhibits remarkable new features, so one 
question that comes out naturally is whether this novel magnetic background could promote some kind of pairing mechanism between electrons moving 
on top of such magnetic profile.  As a first step in this direction, we identify  and study in detail a simple low-energy effective 
description that reproduces the correct spin phenomenology and that could serve as a first step to analyze the coupling between localised spins and 
conduction-electron spin which could, in turn, give rise to interesting electron transport phenomena\cite{Zang_2011}.  For this purpose, based in a combined analysis using
a variational approach and large-scale Monte Carlo simulations, we get quantitative predictions for the existence, the location
and the sizes of the AF-SkX phase induced by a external magnetic field.

The rest of the paper is organised as follows. In Sec. \ref{sec-Hamiltonian} we present the microscopic Hamiltonian and construct the continuous low-energy description. In  Sec. \ref{sec-ansatz} we propose variational Ans\"{a}tze for the different phases that we expect, from the numerical simulation results\cite{Rosales_15}. In Sec. \ref{sec-results} we present the phase diagram of the continuous model obtained with these variational Ans\"{a}tze. We find a rich low temperature behavior of the system as the magnetic field is varied, recovering all the previously observed phases.
The system goes from a helical phase (HL) at low fields to an antiferromagnetic skyrmion lattice phase (AF-SkX) for larger values of the field and then, before the ferromagnetic saturated phase (FM), there seems to be an intermediate phase, which we call sublattice-uniform (SU) phase, that is described below. All our analytical predictions are supported by Monte Carlo (MC) simulations of the microscopic Hamiltonian. We conclude in Sec. \ref{sec-conclusion}  with a summary and discussion of our results.

%

%%%%%%%%%%%%%%%%%%%%%%%%%%%%%%%%%%%%%%%%%%%%%%%%%%%%%%%%%%%%%%%%%%%%%%%%%%%%%%%%
\section{Microscopic Hamiltonian and continuous limit}
\label{sec-Hamiltonian}
%%%%%%%%%%%%%%%%%%%%%%%%%%%%%%%%%%%%%%%%%%%%%%%%%%%%%%%%%%%%%%%%%%%%%%%%%%%%%%%%

\begin{figure}[!]
\includegraphics[width=6.5cm]{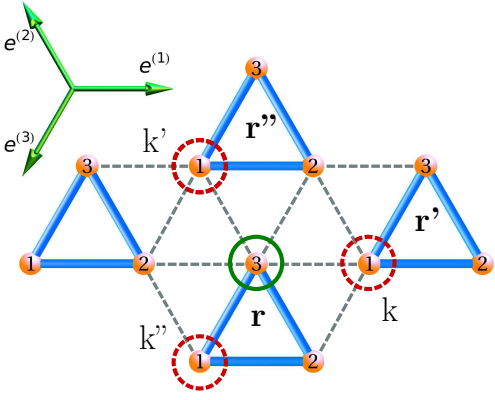}
\caption{Triangular lattice: $\rv,\rv',\rv''$ indicate the plaquettes involved in a given term of the Hamiltonian density. As an example, the sublattice ``3'' (inside the green circle) in the plaquette in $\rv$ has three first neighbors ``1'' (each one inside a red dashed circle) indicated by labels k, k' and k''. In top-left is shown the bond directors vectors.}
\label{fig:lattice}
\end{figure}

We begin with the classical spin Hamiltonian in the triangular lattice (Fig. \ref{fig:lattice})  given by
\be
\mathcal{H}=\sum_{<\rv\rv'>}\left[J\, \Sp_{\rv} \cdot \Sp_{\rv'} + \Dp_{\rv\rv'} \cdot \left(\Sp_{\rv} \times \Sp_{\rv'}\right)\right] -\Bv \cdot\sum_{\rv} \Sp_{\rv},
\label{eq:Hamiltonian}
\ee
where $J>0$ is the antiferromagnetic exchange constant, vectors $\Dp_{\rv\rv'}$ describe the antisiymmetric DM interaction ($\Dp_{\rv\rv'}\equiv -\Dp_{\rv'\rv}$) that stabilizes
the AF-SkX phase recently described in Ref. 31 under an external magnetic field  $\Bv=B\,\hat{z}$
and $<\rv\rv'>$ indicates nearest neighbors (NN).

With the aim to obtain the continuous limit, it is more convenient to rewrite the previous Hamiltonian as a sum of plaquette Hamiltonians $\mathcal{H}=\sum_{\rv}H_{\rv}$, where $\rv$ is the plaquette label. This procedure allows us to write the Hamiltonian density $H_{\rv}$ in a symmetric way in terms of two indices $i,j$ which denote the sublattices ``$i$'' and ``$j$'' (that from now on will be called flavor index), and an index $(k)$ denoting which neighbor of the sublattice ``$j$'' we are considering\cite{DR_89} (see Fig. \ref{fig:lattice}). The indices $i, j$ and $k$ run from 1 to 3. From now on the $\mathbf{r}$ dependence of $H_{\rv}$ (and the terms included in $H_{\rv}$) and in the spin variables $\Sp_{j}$ is suppressed to simplify the notation, i.e. $H_{\rv}\to H$, $\Sp_i(\rv)\to \Sp_i$.
The plaquette Hamiltonian density $H$  reads
\bea
H&=&H_E+H_{DM}+H_z\nonumber\\
H_E&=& \frac J 6  \sum_i \sum_{j\neq i}\sum_k \Sp_i \cdot \Sp_j^{(k)}\nonumber\\
H_{DM}&=&\frac 1 6 \sum_{i}\sum_{j\neq i}\sum_{k} \Dp^{(k)}_{ij} . \left(\Sp_i\times \Sp^{(k)}_j\right)\nonumber\\
H_Z&=&-\frac{1}{3}\sum_{i} \Bv \cdot \Sp_{i}.
\label{eq:density-H}
\eea

Assuming that each spin flavor varies slowly, an approximation that holds both near the ferromagnetic and the antiferromagnetic
order, we can describe the continuum limit of each spin flavor by a smooth field configuration. Under such an assumption we can expand the value of the spin field $\Sp_j^{(k)}$ at site $j$ around the position of the spin $\Sp_i$ as follows:
\bea
\Sp^k_j &=&\Sp_j + a\left[\ev^{(k)}_{ij} . \nabla \right] \Sp_j  + \frac{a^2}{2} \left[\ev^{(k)}_{ij} . \nabla \right]^2 \Sp_j+\mathcal{O}(a^3)\nonumber\\
\label{eq:expansionI}
\eea
where $a$ is the nearest-neighbor distance, $\ev^{(k)}_{ij}= -\text{sgn}[P(ij)]\ev^{(k)}$, where $P(ij)$ is the permutation $(123) \rightarrow (ijl)$, with $\ev^{(1)}=(1,0)$, $\ev^{(2)}=(-\frac{1}{2},\frac{\sqrt{3}}{2})$, $\ev^{(3)}=(-\frac{1}{2},-\frac{\sqrt{3}}{2})$
the bond directors (see Figure \ref{fig:lattice}).

Performing a gradient expansion the exchange Hamiltonian density up to second order in $a$ reads:
\bea
H_E&=& J  \sum_i \sum_{j\neq i}\left(\frac{a^2}{8}\Sp_i \nabla^2\Sp_j-\frac{1}{2}\Sp_i\cdot \Sp_j\right)+\text{const.},
\label{eq:He}
\eea

The next term in Eq. (\ref{eq:density-H}) corresponds to the DM Hamiltonian density $H_{DM}$.
Let us define a cyclic DM vectors $\Dp^{(k)}_{ij}=D\,\; \ev^{(k)}_{ij} $ as in Ref. 31. Using the gradient expansion (\ref{eq:expansionI}), $H_{DM}$,  up to second order in $a$, becomes:
\begin{widetext}
\bea
H_{DM}&=&\frac{1}{6}\sum_{i}\sum_{j\neq i}\sum_{k}\Big[ \Dp^{(k)}_{ij} . \left(\Sp_i\times \Sp_j\right)+a\,\Dp^{(k)}_{ij}.(\Sp_i\times (\ev^{(k)}_{ij} . \nabla) \Sp_j)
+ \frac{a^2}{2}\Dp^{(k)}_{ij} . (\Sp_i\times(\ev^{(k)}_{ij} . \nabla )^2\Sp_j)  \Big]
\label{eq:expDMI}
\eea
\end{widetext}

The first term on the right side in (\ref{eq:expDMI}) vanishes, because $\sum_{k}\ev^{(k)}_{ij}=0$. Using the definitions of $\Dp^{(k)}_{ij}$ and $\ev^{(k)}_{ij}$, the second term reads
\bea
&&\frac{a}{6} \sum_{i}\sum_{j\neq i}\sum_{k} \Dp^{(k)}_{ij} \cdot\left[\Sp_i\times (\ev^{(k)}_{ij} \cdot\nabla ) \Sp_j\right]\nonumber\\
&=&-\frac{ aD\,}{4} \sum_{i}\sum_{j\neq i} \Sp_i \cdot(\nabla \times \Sp_j)\nonumber
\eea

Finally, the last term in (\ref{eq:expDMI})  vanishes due to the antisymmetry of the DM-coupling ($\Dp_{\rv\rv'}\equiv -\Dp_{\rv'\rv}$).
Hence the complete DM Hamiltonian density reads:
\bea
H_{DM}&=&-\frac{aD\,}{4} \sum_i \sum_{j\neq i} \Sp_i \cdot(\nabla \times \Sp_j)
\label{eq:Hdm}
\label{eq}
\eea
Putting all the pieces together we can write the complete Hamiltonian density ($H$) for an antiferromagnetic triangular chiral magnet in the continuous limit as
\bea
H&=&J \sum_{i,j\neq i}\frac{1}{2}\Sp_i\cdot \Sp_j+\frac{a^2}{8}\Sp_i \nabla^2\Sp_j-\frac{aD\,}{4J} \Sp_i \cdot(\nabla \times \Sp_j)\nonumber\\
&&-\frac{1}{3}\sum_{i}\Bv \cdot\Sp_i
\label{eq:density-Hamiltonian}
\eea

The equations of motion of the previous Hamiltonian are non-linear and fairly difficult
to solve analytically. Instead we study the Hamiltonian density proposing
different families of Ans\"atze. In order to gain some
intuition on the possible expressions we rewrite (\ref{eq:density-Hamiltonian}) by introducing a
non-independent variable $M=\sum_i \Sp_i$, the plaquette magnetization. After some trivial algebraic manipulations Eq. (\ref{eq:density-Hamiltonian}) can be recasted in the following form:
\bea
\label{eq:density-Hamiltonian-M}
H&=&H_M+\sum_{i=1}^{3}H_i\\
H_M&=&\frac J 2(\Mv^2 -3)+ a^2 \frac J 8 \Mv \cdot \nabla^2 \Mv -\frac{ aD\,}{4} \Mv \cdot (\nabla \times \Mv)\nonumber\\
H_i&=& -a^2 \frac J 8  \Sp_i \nabla^2 \Sp_i  + \frac{aD\,}{4} \Sp_i \cdot(\nabla \times \Sp_i)- \frac{1}{3} \Bv \cdot \Sp_i\nonumber.
\eea

Some remarks are in order: we notice that the Hamiltonian density has
been separated in four pieces. The first piece corresponds to a
Hamiltonian density $H_M$ for the plaquette magnetization, while the rest corresponds 
to three copies of the same Hamiltonian density $H_i$, one for each
flavor. Each of these $H_i$ has exactly the form of the
{\it ferromagnetic} non-linear sigma model studied by \textcite{HZY_10}
for chiral magnets. This is a crucial observation that, together with the 
knowledge of the finite temperature phases of the
system \cite{Rosales_15}, motivate the Ans\"atze that we propose in the
following section. We also call the attention to the
derivative term in the magnetization density that, at first sight,
seems to  lead to an energy unbounded
from below. This is just an artifact of the introduction of the
non-independent variable $\Mv$. The Laplacian term
in the magnetization density has its origin in the exchange interaction
term
\bea
&J&\sum_i \sum_{j\neq i}\Sp_i\cdot \Sp_j\nonumber\\
&=&\frac J 2\left(\Mv^2 -3\right)+ a^2 \frac J 8  \Mv \cdot \nabla^2 \Mv - a^2 \frac J 8  \sum_i  \Sp_i \nabla^2 \Sp_i,\nonumber\\
\label{eq:He}
\eea
and since the left-hand side of Eq.(\ref{eq:He}) is bounded
from below, the right-hand side should be so as well. This means that the eventual large contribution 
that could arise from the term $a^2 \frac J 8 \Mv \cdot \nabla^2 \Mv$ will be compensated by the term $-a^2 \frac J
8 \sum_i \Sp_i \nabla^2 \Sp_i$. Hence, the
full Hamiltonian remains bounded from below, as the original
Hamiltonian. In fact, as it will be explicitly described in the next section, the
derivative terms of the magnetization on the solutions are orders of
magnitude smaller than the rest of the terms that appear in the
Hamiltonian density (Eq. \ref{eq:density-Hamiltonian-M})

%%%%%%%%%%%%%%%%%%%%%%%%%%%%%%%%%%%%%%%%%%%%%%%%%%%%%%%%%%%%%%%%%%%%%%%%%%%%%%%%
\section{Ans\"atze and Effective Low Energy Hamiltonian}
\label{sec-ansatz}
%%%%%%%%%%%%%%%%%%%%%%%%%%%%%%%%%%%%%%%%%%%%%%%%%%%%%%%%%%%%%%%%%%%%%%%%%%%%%%%%
The possibility to rewrite the continuum Hamiltonian as a sum of
flavour Hamiltonian densities ($H_i$) plus a plaquette
magnetization contribution ($H_M$), allows for an
intuitive analysis. We mentioned in Sec. \ref{sec-Hamiltonian} that flavour
Hamiltonians are exactly the continuum model found by Bogdanov and
collaborators for ferromagnetic chiral magnets \cite{BY_89,BH_94}. In
\textcite{HZY_10}, the authors have shown that this
Hamiltonian admits a non-trivial periodic magnetic texture known as
skyrmion-lattice (SkX), {\it i.e.} (a periodic arrange of
skyrmions). So, the presence of three independent $H_i$
Hamiltonians in the continuum limit strongly suggests the possibility 
of  the same kind of non-trivial SkX solutions on each sublattice. 

These three independent equivalent SkX solutions need to be arranged in such a way that their sum, $\Mv$,
minimizes the corresponding magnetization Hamiltonian. 

%%%%%%%%%%%%%%%%%%%%%%%%%%%%%%%%%%%%%%%%
\subsection{Skyrmion crystal Ansatz}
%%%%%%%%%%%%%%%%%%%%%%%%%%%%%%%%%%%%%%%%

The proposed approximate solution to one spin flavour Hamiltonian can be
constructed as a superposition of three helical solutions with wave
vectors $\mathbf{k}_{\mu}$ satisfying $\sum_{\mu}\mathbf{k}_{\mu}=\bf{0}$ ($\mu=1,2,3$) in the plane of the sample with relative angles of $2\pi/3$
\cite{OCK_12}. The approximate skyrmion lattice solution then reads:
\begin{widetext}
\bea
\mathbf{n}_{SkX}(\rv)&=&\frac{1}{n} I_{xy}\sum_{\mu}\sin{[\frac{2\pi}{T}\mathbf{k}_{\mu}\cdot\mathbf{r}+\theta_{\mu}]}\mathbf{e}_{xy,\mu}+(m_{z}+I_{z}\sum_{\mu}\cos{[\frac{2\pi}{T}\mathbf{k}_{\mu}\cdot\mathbf{r}+\theta_{\mu}]})\mathbf{e}_{z},
\label{eq:ansatz-nSkX}
\eea
\end{widetext}
where $T$ is the period of each helix; $n$ fixes the appropriate normalization $|\mathbf{n}_{SkX}|=1$ which restricts the values of the amplitudes $I_{xy}$ (in-plane) and $I_z$ (perpendicular to the xy-plane) and the homogeneous contribution to the magnetization in the  $z-$direction $m_z$. $\mathbf{e}_{xy,\mu}$ are arbitrary unit vectors lying on the xy-plane 
satisfying $\sum_{\mu}\mathbf{e}_{xy,\mu}=\bf{0}$, while the phases $\theta_{\mu}$ satisfy $\cos(\theta_1+\theta_2+\theta_3)=-1$\cite{OCK_12}.

The helix period $T$, that becomes the skyrmion lattice
parameter, can be determined as a function of $I_{xy}, I_z$ and $m_z$ by
energy scale analysis (see the Appendix \ref{appESA}). 
Now, the proposed Ansatz for the full solution reads:
\bea
\Sp_1(\rv)&=&\mathbf{n}_{SkX}(\rv),\;\;\Sp_2(\rv)=\mathbf{n}_{SkX}(\rv+{\bf T}_1),\nonumber\\ \Sp_3(\rv)&=&\mathbf{n}_{SkX}(\rv+{\bf T}_2),
\label{eq:ansatz-S1yS2yS3}
\eea
where ${\bf T}_1, {\bf T}_2$ are arbitrary translations in the xy-plane.

%%%%%%%%%%%%%%%%%%%%%%%%%%%%%%%%%%%%%%%%
\subsection{Helical Ans\"atz}
%%%%%%%%%%%%%%%%%%%%%%%%%%%%%%%%%%%%%%%%

In the helical phase, the spin structure is a special case of the Ansatz (\ref{eq:ansatz-nSkX}) and  consists of three interpenetrating spirals on each sublattice, as in Eq. (\ref{eq:ansatz-S1yS2yS3}), but with a single-$\mathbf{k}_{\mu_0}$ mode
\begin{widetext}
\bea
\mathbf{n}_{H}(\rv)&=&\frac{1}{n} I_{xy}\sin{[\frac{2\pi}{T}\mathbf{k}_{\mu_0}\cdot\mathbf{r}+\theta]}\mathbf{e}_{xy}+(m_{z}+I_{z}\cos{[\frac{2\pi}{T}\mathbf{k}_{\mu_0}\cdot\mathbf{r}+\theta]})\mathbf{e}_{z},
\label{eq:ansatz-H}
\eea
\end{widetext}
where again the constant $n$ fixes the normalization $|\mathbf{n}_{H}|=1$.
%%%%%%%%%%%%%%%%%%%%%%%%%%%%%%%%%%%%%%%%
\subsection{Uniform sublattice Ansatz}
%%%%%%%%%%%%%%%%%%%%%%%%%%%%%%%%%%%%%%%%

The magnetic phase diagram for the model defined by Eq. (\ref{eq:Hamiltonian}) with $D=0$ has been discussed in \cite{Triangular1,Triangular2}. At zero temperture and zero magnetic field the ground state is a
planar configuration with spins arranged in a $120^{\circ}$ structure described by the wave vector $k= (4\pi/3, 0)$. In a magnetic field the energy is minimized when the constraint

\be
\mathbf{S}_{1}+\mathbf{S}_{2}+\mathbf{S}_{3}=\mathbf{B}/(3J),
\label{lab:const}
\ee
is fulfilled on each plaquette.
This constraint persists up to the saturation field $B=9J$, where the spins are fully polarized.

For $D\neq 0$ the previous discussion breaks down since the DM term stabilizes new configurations.
However, it is worth noting that even for $D\neq 0$ there exist spin configurations in which the DM contribution cancels out. This is the case when the spin field on each sublattice is uniform. This is easily seen from our effective model, since the DM term contains derivatives of the spin fields.
If one goes back to the microscopic model, one can show that the sum of the interactions (through DM)
of a specific spin with its six neighbors is zero for the present choice of the $\mathbf{D}$ vectors.
Thus, for this kind of configurations, which we call SU for ``sublattice uniform'' from now on, the constraint given by Eq. (\ref{lab:const}) is still valid, and this is an equilibrium state to be considered in the following discussion of the phase diagram.

The energy per plaquette of the states satisfying the constraint (\ref{lab:const}) is field dependent, independent of $D$ and is given by:
\be
E_{SU}=-\frac{B^{2}}{18J}-J\frac{3}{2}.
\ee
Finally, at the saturation the energy per plaquette of the ferromagnetic state (for $B>9J$) is
\be
E_{FM}=3J-B.
\ee

Now that we have described the Ans\"atze under which we will study
the Hamiltonian, we are in the position to compare the values of the terms
that include derivatives of $\Mv$, to the rest of the terms included
in the Hamiltonian density (\ref{eq:density-Hamiltonian-M}).

\begin{figure}[!]
\includegraphics[width=7.5cm]{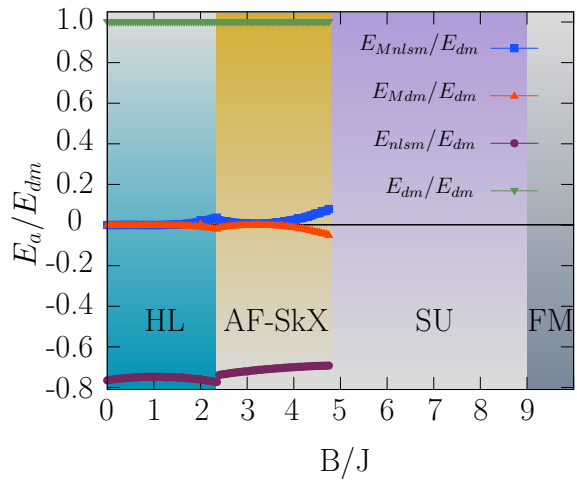}
\caption{(Color online): Comparison between different contibutions from the density Hamiltonian (\ref{eq:density-Hamiltonian-M}) for the case  $D/J=1/2$. We compare the four terms $E_a=\{E_{nlsm},E_{dm},E_{Mnlsm},E_{Mdm}\}$ (see Eqs. (\ref{eq:Enlsm})-(\ref{eq:EMdm}) ) using  $E_{dm}$  as scale.  In the HL and AF-SkX phases, the dominant terms are those coming from $\sum_i H_i$ in Eq. (\ref{eq:density-Hamiltonian-M}). All these terms are zero in the SU and FM (homogeneous) phases.}
\label{fig:TermsEvsB}
\end{figure}

First, let us analyze these terms in the Helix phase. In this case, the plaquette magnetization 
corresponds to a superposition of three helical waves, each one given by Eq. (\ref{eq:ansatz-H}), separated (in space) by a 
translation in the direction of propagation. In the case where the distance between peaks is uniform (i.e. the phase difference of each cosine is $2\pi/3$)
it is straightforward to see from the Ansatz (\ref{eq:ansatz-H}) that $\Mv$ will show small spatial variations: $\Mv(\rv)\approx\Mv=\text{const.}$. For the SkX phase, a similar analysis  drives to the same conclusion. These statements are confirmed by our numerical calculations performed for both Ans\"atze for different values of the coupling $D$ and
as a function of $\Bv$. Our results show that the plaquette magnetization 
is almost constant leading to the conclusion that the contribution of the laplacian and curl terms in $H_M$,
are two orders of magnitude smaller than the rest of the terms present
in the Hamiltonian density (\ref{eq:density-Hamiltonian-M}) (see figure \ref{fig:TermsEvsB}). To this purpose we compare the four contributions (with spatial derivatives) of the total energy, namely: $E_{nlsm}$, $E_{dm}$ , $E_{Mnlsm}$  and $E_{Mdm}$, where 
\bea
\label{eq:Enlsm}
E_{nlsm}&=&-a^2 \frac J 8\sum_{\rv}\sum_i\Sp_i \nabla^2 \Sp_i \\
\label{eq:Edm}
E_{dm}&=&\frac{aD\,}{4}\sum_{\rv}\sum_i \Sp_i \cdot(\nabla \times \Sp_i) \\
\label{eq:EMnlsm}
E_{Mnlsm}&=&a^2 \frac J 8\sum_{\rv}\Mv \cdot \nabla^2 \Mv  \\
\label{eq:EMdm}
E_{Mdm}&=&-\frac{ aD\,}{4}\sum_{\rv} \Mv \cdot (\nabla \times \Mv) 
\eea

In figure \ref{fig:TermsEvsB} we plot the ratios between the four terms (\ref{eq:Enlsm})-(\ref{eq:EMdm})  setting $E_{dm}$ as the scale, for the case $D/J=1/2$. We observe that in the HL and AF-SkX phases both  $E_{Mnlsm}/E_{dm}$ as well as $E_{Mdm}/E_{dm}$ are neglegible in almost all the field range, except for two narrow windows around the transition fields where the value of these ratios are smaller than $5\times10^{-2}$. In the homogeneous SU and FM phases all the terms with derivatives are zero. This behaviour is repeated in all the range that we have explored $D/J <1$, leading to the conclusion that the contributions of the laplacian and curl terms in $H_M$, are at least two orders of magnitude smaller than the rest of the terms present in the Hamiltonian.

Monte Carlo simulations show that the spatial variation of the magnetization is
small compared to the variation of the spin on each sublattice, confirming the
observation made by the variational approach.
Based on the previous analysis, we end up this section by proposing a simplified low-energy effective
Hamiltonian that captures the low-energy physics of the antiferromagnetic chiral magnet given by Eq. (\ref{eq:Hamiltonian}).

%%%%%%%%%%%%%%%%%%%%%%%%%%%%%%%%%%%%%%%%
\subsection{Effective low energy theory}
%%%%%%%%%%%%%%%%%%%%%%%%%%%%%%%%%%%%%%%%

From the previous discussion we can rewrite Eq.
(\ref{eq:density-Hamiltonian}) in the following form:

\bea
\label{eq:density-HamiltonianFinal}
H&=&\sum_{i=1}^{3}H_i+H_M\\
H_i&=& -a^2 \frac J 8  \Sp_i \nabla^2 \Sp_i  + \frac{aD\,}{4} \Sp_i \cdot(\nabla \times \Sp_i)- \frac{1}{3} \Bv \cdot \Sp_i \nonumber \\
H_M&=&\frac J 2(\Mv^2 -3) \nonumber.
\eea

It is remarkable that this continuum effective Hamiltonian can be thought as the sum of three Ginzburg-Landau (GL) effective actions (one for each flavor/sublattice) plus a term $H_{M}$ that couples them. From the first term of the sum one could expect, separately on each sublattice, the well known three phases, HL, SkX and FM. 

%%%%%%%%%%%%%%%%%%%%%%%%%%%%%%%%%%%%%%%%%%%%%%%%%%%%%%%%%%%%%%%%%%%%%%%%%%%%%%%%
\section{Results and Phase Diagram}
\label{sec-results}
%%%%%%%%%%%%%%%%%%%%%%%%%%%%%%%%%%%%%%%%%%%%%%%%%%%%%%%%%%%%%%%%%%%%%%%%%%%%%%%%

In this Section we construct the full phase diagram of the Hamiltonian (\ref{eq:density-HamiltonianFinal}), paying particular attention to the appearance of the topological AF-SkX phase.

In the study of the phase diagram we consider four phases, namely HL and SkX phases with energies $E_{HL}$ and $E_{SkX}$ respectively, together with SU and FM phases presented in section \ref{sec-ansatz}. To find the minimum energy configuration we fix the variational parameters in an self consistent way by using the Nelder-Mead simplex method that is one of the most used for direct optimization\cite{NelderMeadMethod}.
The procedure consists of introducing an initial guess for $I_{xy}$ and $m_z$, and determine variationally the values of $\mathbf{T_1}$ and $\mathbf{T_2}$ self consistently. 

\begin{figure}[ht]
\includegraphics[width=7.2cm]{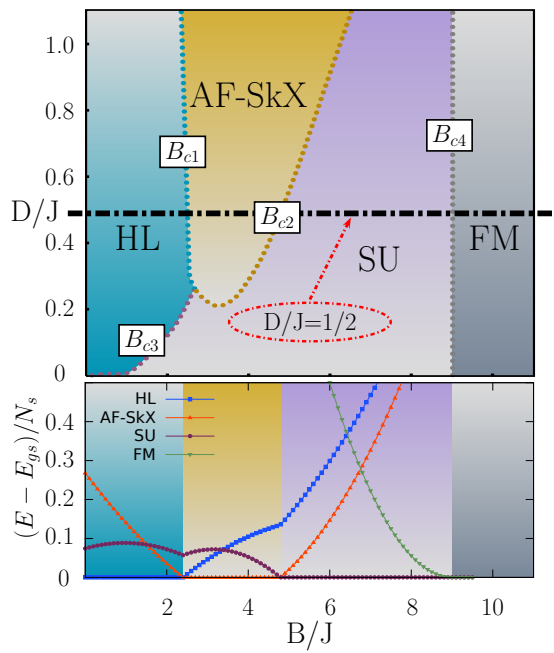}
\caption{(Color online) Top: Phase diagram: dotted lines correspond to the boundaries between different phases labeled by the fiels $B_{c1}...B_{c4}$ (functions of $D$). Bottom: The energies ($E_{\text{phase}}-E_{\text{gs}}$) of the three states of triangular antiferromagnetic chiral magnet with $D/J=1/2$ as a function of the
external magnetic field $B$. This example corresponds to the path indicated by the dashed black line in the phase diagram (Top).}
\label{fig:phasediagram}
\end{figure}

\begin{figure}[!]
   \centering
   \fbox{\includegraphics[width=0.3\textwidth]{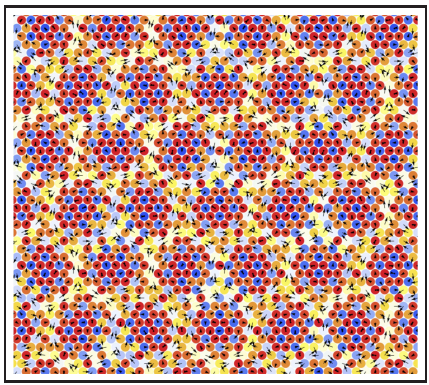}}
  \caption{(Color online)  Representative magnetic texture obtained by low energy effective theory (Eq. (\ref{eq:density-HamiltonianFinal})) and the variational Ans\"atz for $D/J=1/2$ and $B/J=3$.}
   \label{fig:ES1}
\end{figure}

The minimization of the variational energies for the different phases leads to the phase diagram shown in Fig. \ref{fig:phasediagram} (top) where  the boundaries of the phases result from level crossings as shown in the Fig. \ref{fig:phasediagram} (bottom). As an example, in Fig. \ref{fig:ES1} we show a representative spin  texture obtained by the variational Ansatz in the AF-SkX phase ($D/J=1/2$ and $B/J=3$).

The main features of this diagram is the presence of the four phases, namely HL, AF-SkX, SU and FM. in a wide region of $D-B$ ($D>0$) space. However, there exists a critical value $D_{c}\approx0.2$ for the skyrmion lattice to be stable. Below this value,
the skyrmion lattice phase is excluded irrespectively of the magnitude of the external field.
The phase diagram for small fields is dominated by a helical phase with a wave vector lying in the plane.
This phase starts at zero magnetic field $B=0$  and extends to $B_{c3}$ for $D<D_{c}$ and to $B_{c1}$ for $D>D_{c}$ (see Fig. \ref{fig:phasediagram}).

The phase diagram presents a wide region with a complex magnetic texture that is described by the superposition of three skyrmion
lattices, one for each flavor. The region of the parameter space where this phase is stable is delimited by the curves $B_{c1}$ and $B_{c2}$.  From $B_{c2}$
and $B_{c3}$ up to the saturation field $B_{c4}$ the SU phase is realized.

\begin{figure}[!]
\includegraphics[width=7.5cm]{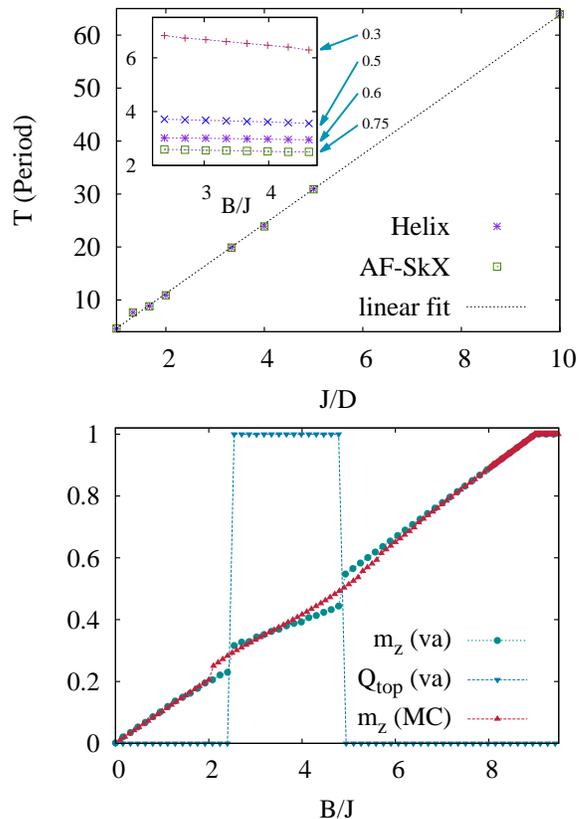}
\caption{Top: Variation of the period with $J/D$ for SkX (green squares) and HL (purples stars). Inset: The skyrmion radius as a function of the external magnetic field for several values of $D$. Bottom: Topological charge $Q_{top}$ and magnetization $m^{T}_{z}$ as a function of the magnetic field for $D=0.5J$. (Bottom): Magnetization vs magnetic field $B$  calculated by variational Ans\"atz and by MC simulations.}
\label{fig:T_and_M}
\end{figure}

For the HL and AF-SkX phases, the optimized value of the period $T$ shows a small linear dependence in the external field (the same for both phases as obtained by MC simulations\cite{Rosales_15}). 
In Fig. \ref{fig:T_and_M} we see that the mean period takes the same values for the HL state and for AF-SkX state as $T(D)\simeq\frac{6.57}{D}-1.95$.

For $D/J=1/2$, we get $T\approx 10.8\pm0.6$ (this value should be compared with the wavelength $T\approx 11.4$
of the HL and AF-SkX phases found in Ref. 31 obtained by numerical simulations of finite-size systems). 
We can define the radius of a skyrmion (in one sub-lattice) as the radius of the circumference of the contour defined by $n_{z}=0$. In the inset of Fig. \ref{fig:T_and_M} (top) we show the skyrmion size as a function of the magnetic field. We observe that the behavior
of the optimal skyrmion spacing as a function of the magnetic field varies very slowly in the region of the AF-SkX phase due to its topological stability.  This behavior translates precisely in a wide range of stability of the AF-SkX phase in which the skyrmion number is fixed.

In order to capture the topological character of the field configuration for each spin flavor we introduce the topological index $Q_{top}$ and define the total (normalized) magnetization (z-component):

\bea
Q_{top}&=&\frac{1}{4\pi}\int_{u.c.}\mathbf{n}\cdot(\partial_{x}\mathbf{n}\times\partial_{y}\mathbf{n})\,d^{2}r,\\
m^{T}_{z}&=&\frac{1}{A_{u.c.}}\int_{u.c.}n_z\,d^{2}r,
\eea
where the integration is performed in a unit cell of the magnetic texture with area $A_{u.c.}$ (see Appendix \ref{appESA}).

In Fig. \ref{fig:T_and_M} (bottom) we show the behavior of the magnetization and the topological charge as a function of the magnetic field. We see that the helical phase corresponds to a trivial configuration with $Q_{top}=0$ whereas in the SkX phase (triple-helix state) $Q_{top}=1$ because each unit cell contains only one skyrmion. The magnetization curve reveals an almost linear growth up to the saturation field.  However, we see two discontinuities suggesting a first order phase transition form HL to  AF-SkX phase and from AF-SkX to SU phase.

\begin{figure}[!]
\includegraphics[width=8.5cm]{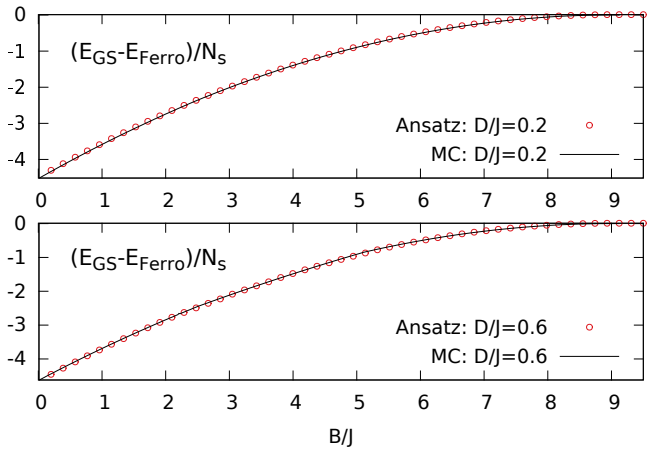}
\caption{(Color online): Comparison of the energies vs magnetic field ($B$) obtained by variational Ans\"atz and MC simulations for $D/J=0.2$ and $0.6$. The specific values of $D/J$ were chosen as representative of two possible paths as a function of the external field $B$: one that goes directly from the HL phase to SU phase  ($0.2$ case); and another in which the path goes through the AF-SkX phase ($0.6$). }
\label{fig:EvsB_Va_and_MC}
\end{figure}

In order to confirm the results from the variational analysis, we numerically examine the ground state of the model (\ref{eq:Hamiltonian}) by Monte Carlo simulations based on the standard heatbath method combined with the over-relaxation method. We have implemented periodic boundary conditions for $N=3600$ sites. A run at each magnetic field or temperature contains typically $0.1-1\cdot10^6$  Monte Carlo steps (MCS's) for initial relaxation and twice MCS's during the calculation of mean values. In Fig. \ref{fig:T_and_M} (bottom) we compare magnetization vs magnetic field for the  minimized variational solution and by MC simulations for $D/J=1/2$. We observe qualitative agreement between both methods. However, the behaviour of the magnetization differs when the system switches from one phase to another.  This may be due to finite size effects of the MC simulations and that in the transition region, the variational solution does not include higher order modes in ${\bf k}$. In Fig. \ref{fig:EvsB_Va_and_MC} we compare  the ground state energy as a function of the magnetic field obtained from the minimization of the variational energies for the different phases and by MC simulations for two values of $D/J$. The excellent agreement between both results further supports the variational analysis of the continuous limit of the microscopic Hamiltonian given by Eq. (\ref{eq:density-HamiltonianFinal}).

%

%%%%%%%%%%%%%%%%%%%%%%%%%%%%%%%%%%%%%%%%
\section{Discussion and Conclusion }
\label{sec-conclusion}
%%%%%%%%%%%%%%%%%%%%%%%%%%%%%%%%%%%%%%%%

To summarize, we have constructed a low-energy theory describing the behavior of the Heisenberg model in the triangular lattice including Dzyaloshinskii-Moriya interactions and the magnetic field. Our low energy effective theory given in Eq. (\ref{eq:density-HamiltonianFinal}), notwithstanding its simplicity, displays a plethora of phenomena of current interest in the context of topological magnetic phases. The effective theory obtained surprisingly consists of three independent Hamiltonian densities ($H_i$) similar to those found by  Bogdanov {\it et al.}\cite{BY_89,BH_94} and Nagaosa {et al.}\textcite{HZY_10} in the context of ferromagnetic systems. Each one of these admit non-trivial  magnetic structures known as skyrmion-lattices (SkX). In addition to these terms, there is a plaquette magnetization contribution ($H_M$) which couples the previous $H_i$'s. The  low-energy  theory predicts a AF-SkX crystal  phase which consists of three interpenetrating SkX states as observed in numerical Monte Carlo simulations\cite{Rosales_15}. The low-energy effective Hamiltonian reproduces the correct spin phenomenology and could serve as a first step to analyze the coupling to charge degrees of freedom. In addition we numerically examined the ground state of the micropcopic model by Monte Carlo simulations showing a very good agreement between both methods. Finally, the remarkable stability that presents the  AF-SkX phase for a wide range of magnetic fields can have interesting consequences in the context of the anomalous Hall effect.

%%%%%%%%%%%%%%%%%%%%%%%%%%%%%%%%%%%%%%%%%%%%%%%%%
\section*{Acknowledgments}
%%%%%%%%%%%%%%%%%%%%%%%%%%%%%%%%%%%%%%%%%%%%%%%%%
The authors specially thank Pierre Pujol, Nicol\'as Grandi and Gerardo Rossini for fruitful discussions. This work was partially supported by CONICET (PIP 0747) and ANPCyT
(PICT 2012-1724).

%%%%%%%%%%%%%%%%%%%%%%%%%%%%%%%%%%%%%%%%%%%%%%%%%%%%%%%%%%%%%%%%%%%%%%%%%%%%%%%%
%%%%%%%%%%%%%%%%%%%%%%%%%%%%%%%%%%%%%%%%%%%%%%%%%%%%%%%%%%%%%%%%%%%%%%%%%%%%%%%%
\appendix

%%%%%%%%%%%%%%%%%%%%%%%%%%%%%%%%%%%%%%%%%%%%%%%%%%%%%%%%%%%%%%%%%%%%%%%%%%%%%%%%
\section{Energy Scale Analysis}
\label{appESA}
%%%%%%%%%%%%%%%%%%%%%%%%%%%%%%%%%%%%%%%%%%%%%%%%%%%%%%%%%%%%%%%%%%%%%%%%%%%%%%%%

The magnetic textures considered in section \ref{sec-ansatz}, namely helix and AF-SkX, are periodic configurations in $x$ and $y$ directions with periods $\alpha\,T$ and 
$\beta\,T$ respectively, with $\alpha,\beta$  to be fixed by the symmetry of the texture (for helix $\alpha=\beta=1$, and for AF-SkX $\alpha=1$ and $\beta=2/\sqrt{3}$).
This allows to calculate the total energy as the energy of a cell (of area $A_{u.c.}=\alpha\beta T^2$) times the number of cells, $L^2/(\alpha\beta T^2)$, in the sample. 
In addition, we separate different contributions in the energy density according to the order of spatial derivatives. 
With all this, the total energy can be written as
\be
E(T,I_{xy},I_{z},m_{z})=\frac{L^2}{\alpha\beta T^2}\sum_{i=0}^{2}E_i(T,I_{xy},I_{z},m_{z}),
\ee
with 
\bea
E_{i}(T,I_{xy},I_z,m_{z})&=&\int_{0}^{\alpha T} dx \int_{0}^{\beta T} dy\, \mathcal{E}_{i}(T,I_{xy},I_z,m_{z}),\nonumber\\
\eea
and  $\mathcal{E}_{i}(T,I_{xy},I_z,m_{z})$ denotes the energy density containing ith-order derivatives. We can rewrite the different terms using their properties under scale transformations ($\mathbf{r}\to\mathbf{r}'=\mathbf{r}/T$). We can separate the dependence in $T$ as

\bea
E_{i}(T,I_{xy},I_z,m_{z})&=&\int_{0}^{\alpha T} dx \int_{0}^{\beta T} dy\, \mathcal{E}_{i}(T,I_{xy},I_z,m_{z})\nonumber\\
&=&T^{2-i}\int_{0}^{\alpha} dx' \int_{0}^{\beta} dy'\, \mathcal{E}_{i}(1,I_{xy},I_z,m_{z})\nonumber\\
&=&T^{2-i}E_{i}(1,I_{xy},I_z,m_{z}),\nonumber
\eea
and write the energy of the sample as

\be
E(T,I_{xy},I_z,m_{z})=\frac{L^2}{\alpha\beta}\left[\frac{E_{2}}{T^2}+\frac{E_{1}}{T}+E_{0}\right]\nonumber.
\label{edet}
\ee
This shows that all the dependence in the variable $T$ can be cast as power law prefactors.
\bibliography{references.bib}

\end{document}